\documentclass[aps,pra,superscriptaddress,twocolumn]{revtex4}

\draft 

\usepackage{physics}
\usepackage{graphicx}
\usepackage{dcolumn}
\usepackage{bm}

\usepackage{tikz}
\usetikzlibrary{arrows.meta}

\begin{document}

\title{Ultrashort laser pulse driven currents in conductors: \\physical mechanisms and time scales}

\author{Istv\'{a}n Magashegyi}
\affiliation{ 
Department of Theoretical Physics, University of Szeged, Tisza Lajos k\"{o}r\'{u}t 84, H-6720 Szeged, Hungary
}

\author{L\'{o}r\'{a}nt Zs.~Szab\'{o}}
\affiliation{ 
Department of Theoretical Physics, University of Szeged, Tisza Lajos k\"{o}r\'{u}t 84, H-6720 Szeged, Hungary
}

\author{P\'{e}ter F\"{o}ldi}
\affiliation{ 
Department of Theoretical Physics, University of Szeged, Tisza Lajos k\"{o}r\'{u}t 84, H-6720 Szeged, Hungary
}
\affiliation{ 
ELI-ALPS, ELI-HU Non-profit Ltd., Dugonics t\'{e}r 13, H-6720 Szeged, Hungary
}



\begin{abstract}
The response of conduction band electrons to a local, pulse-like external excitation is investigated. The charge density wave packets that emerge as a consequence of the excitation leave the interaction region with a speed close to the initial state's band velocity, but there are also oscillations with essentially the same frequency as that of the laser field. As a good estimation, the excitation can also be considered as a localized, time-dependent ponderomotive potential, leading to slowly varying current oscillations. The role of all these effects are investigated for different electron energies, carrier frequencies and sizes of the interaction area.
\end{abstract}

\maketitle

\section{Introduction}
Ultrashort bursts of electromagnetic radiation mean a versatile, efficient optical tool that can control and even visualize processes on the femtosecond (fs) timescale \cite{Baltuska2003,U07,Wirth2011,K14}.  This approach is complementary to the usage of frequency stabilized, many-cycle sources, that can e.g., transfer the system from its ground state to a well-defined excited state. For short, wideband pulses the mere time scale and the possibility of tailoring the pulses \cite{Wirth2011} provide a wide variety of promising possible applications, e.g., light-pulse control of electrons in solids (fast, “lightwave electronics” \cite{K14}). Along this line, the time scale of the electronic response of a solid to an ultrashort excitation is crucial.

The appearance of light-induced currents in a dielectric has been demonstrated in Ref.~\cite{Schiffrin2013} using fused silica targets. Theoretical models \cite{DR11,FBY13,WL14} (partially related to high-order harmonic generation \cite{FK96,VMcOKCB14,HIY15}), as well as measurements \cite{SRPSWGPBPYNL14} indicate that these currents are not only being generated on the fs timescale, but they also disappear similarly fast, which is ideal for ultrafast switching. However, the generation of measurable currents in a wide (several times the photon energy) bandgap material requires intense pulses, setting a considerable limitation for practical applications. Conductors, on the other hand, are known to have charge carriers that can produce currents already for weak external fields. This is clearly a more conventional effect, and the primary application of light-induced currents in metals may not be fast switching. Instead, since these currents are easily generated and their time integral (i.e., the charge transferred by the laser pulse) can also be easily measured, the phenomenon can be used for detection purposes. This would lead to affordable all-solid-state devices that can measure the properties of the exciting pulses.

For the sake of simplicity, in the following we consider a one-dimensional model, which, however, can adequately describe the interaction of laser pulses and metallic or semiconductor nanowires \cite{CXR07,LVC03,KSM08} or conducting carbon nanotubes \cite{R10}. (Additionally, effects related to the penetration depth and screening are expected to play a minor role for nanoscale conductors.) The dynamics of the excited electrons is calculated using the time-dependent version of the non-equilibrium Green's function method (TDNEGF) \cite{WJM93,Z05,G14,F15}, which is a standard tool based on the Landauer-B\"{u}ttiker \cite{L88,B86} formalism of transport processes in solids \cite{D95}. This allows us to compute the transmitted current for different energy eigenstates, and finally add these contributions. Using effective mass approximation, in the current work we focus on how fast different initial plane wave states respond to the excitation of the laser pulse.

In order to see the physical mechanisms beyond the numerically calculated dynamics, first we consider switch-on effects, i.e., assume a suddenly increasing, local potential that remains constant after being switched on. Clearly, this should lead to a steady state solution, when the transmission probability does not change in time. The dynamics of this process is visualized using space and time dependent electron densities, which clearly show the way the localized potential pushes the electrons outside the interaction region.

In the second part of the paper, we consider pulsed laser fields as the source of excitation, and show that the time scales that influence the dynamics are determined by i) the oscillation period of the laser field, ii) $\hbar/E(k)$, where $E(k)$ is the energy eigenvalue corresponding to the initial state and iii) the slow oscillations that result from the time-dependent ponderomotive potential related to the envelope of the laser pulse. We analyze in detail how the ratio of these time scales and the importance of their role in the time evolution depend on the physical parameters.

The current paper is organized as follows: We summarize the physical model and the methods to be used in Sec.~\ref{modelsec}. The results will be discussed in Sec.~\ref{resultsec}, first by considering switch-on effects (Subsec.~\ref{lorisubsec}) which is followed by the analysis of laser-induced currents in Subsec.~\ref{istvansubsec}. Conclusions will be given in Sec.~\ref{summarysubsec}.

\section{Model}
\label{modelsec}
In our one-dimensional (1D) model we consider three regions in space, as it is shown by Fig.~1. In the second, interaction region, the Hamiltonian describing the dynamics is given by
\begin{equation}
H(x,t)=\frac{1}{2m}(p-e{A(x,t)})^{2}
+e\Phi(x,t), \label{Ham}
\end{equation}
where $e$ denotes the elementary charge, $m$ is the effective mass of the conduction band electron, $p=-i\hbar \frac{\partial}{\partial x}$ is the canonical momentum, and $A$ and $\Phi$ denote the vector and scalar potentials corresponding to the excitation, respectively. (Note that in 1D, $A$ means the only nonzero component of the vector potential, i.e., $A=A_x.$) The actual space and time dependence of the electromagnetic potentials depends on the problem we consider and also on the choice of the electromagnetic gauge. For the investigation of switch-on effects (Subsec.~\ref{lorisubsec}) we use nonzero $\Phi$ but vanishing $A,$ while the laser pulse with electric field strength $F$ in Subsec.~\ref{istvansubsec} will be described using the velocity gauge, i.e., $\Phi$ will be zero and $A(x,t)=-\int_0^t F(x,t') dt'.$ However, in both cases, both $A$ and $\Phi$ will be zero outside the interaction region, i.e., $H=H_0=\frac{p^2}{2m}$ in regions I and III. We also assume that the electromagnetic potentials are zero for $t<0,$ thus for negative values of time, we have free propagation in all the three domains.

The initial state state of the electrons is assumed to be a plane wave,
\begin{equation}
\Psi(x,t)=\Psi_0(x,t)=e^{i\left[kx-\omega(k) t\right]}, \ \ t<0.
\end{equation}
Note that $H_0 \Psi(x,t) = E(k) \Psi(x,t),$ with $E(k)=\frac{\hbar^2 k^2}{2m}=\hbar \omega(k).$ For the sake of definiteness, we choose positive wave numbers $k,$ i.e., the initial states propagate in the positive $x$ direction. For $t>0,$ when the excitation is nonzero, the solutions of the time dependent Schr\"{o}dinger equation governed by the Hamiltonians $H_0$ (in regions I and III) and $H(t)$ (in region II) will be no longer plane waves. The perturbation in region II generates wave packets that propagate generally in both the positive and negative $x$ directions and superimpose on the initial plane wave. Considering that $k>0,$ the complete wave function in region III will be termed as the "transmitted" wave while in region I, we have a superposition of the "incoming" wave $\Psi_0$ and the "reflected" one. The usual probability current density (which is proportional to the charge current density)
\begin{equation}
j(x,t)=\frac{\hbar}{m}\mathrm{Im} \left[\Psi^*(x,t) \frac{\partial}{\partial x} \Psi(x,t)\right]
\end{equation}
can be used to define the transmission probability (valid in region III)
\begin{equation}
T(x,t)=\frac{j(x,t)}{j_0(x,t)}=\frac{j(x,t)m}{\hbar k}
\end{equation}
as the ratio of current densities corresponding to the incoming plane wave $\Psi_0$ and the transmitted one $\Psi,$ which, clearly,  has to be evaluated in region III. (Note that in the equation above we used that the current density corresponding to $\Psi_0$ is constant both in time and space, i.e., $j_0(x,t)=j_0=\hbar k/m.$) For steady state solutions, $T(x,t)$ is also constant, but this is not the case for time dependent problems.

\begin{figure}
\begin{center}
\begin{tikzpicture}


\draw[-{Stealth[length=3mm, width=1.5mm]}] (0,0) -- (7,0) node[anchor=north] {$x$};

\draw	(2,0) node[anchor=north] {$ x_1 $}
		(5,0) node[anchor=north] {$ x_2 $};

\draw	(1,2.3) node{ \footnotesize \bf Region I }
		(3.5,2.3) node{ \footnotesize \bf Region II }
		(6,2.3) node{ \footnotesize \bf Region III };


\draw[-{Stealth[length=3mm, width=1.5mm]}] (0,0) -- (0,3) node[above, midway, color=black, rotate=90] { $\Phi / A$};


\draw[dotted] (2,0) -- (2,1);

\draw[dotted] (5,0) -- (5,1);


\draw[red,very thick] (0.01,.02) -- (2,0.02);

\draw[red,very thick] (5,0.02) -- (6.5,0.02);

\draw[color=red]	(2.9,1.8) node{\tiny $\Phi(x,t)/A(x,t)$ };

\draw[red,thick] plot[smooth] coordinates {(2,0.02) (2.2,0.2)(2.5,1.1) (3.5,1.4)(4,2)(4.5,0.7) (4.8,0.2) (5,0.02)};


\draw[color=blue,Latex-Latex] (0,0.8) -- (2,0.8) node[midway, above]{\tiny Free propagation };

\draw[color=blue] (1,0.5) node{\tiny (incoming};

\draw[color=blue] (1,0.2) node{\tiny +reflection) };

\draw[color=blue,,Latex-Latex] (2,0.8) -- (5,0.8) node[midway, above] {\tiny Scattering };

\draw[color=blue,-Latex] (5,0.8) -- (7,0.8) node[midway, above] {\tiny Free propagation };

\draw[color=blue] (6,0.5) node{\tiny  (transmission) };

\end{tikzpicture}

\caption{The scheme of the geometry that we consider. Localized external fields (described by the scalar and/or vector potentials) interact with the conduction band electrons in the middle of region II. The electrons can propagate freely in regions I and III.}
\label{timedepfig}
\end{center}
\end{figure}
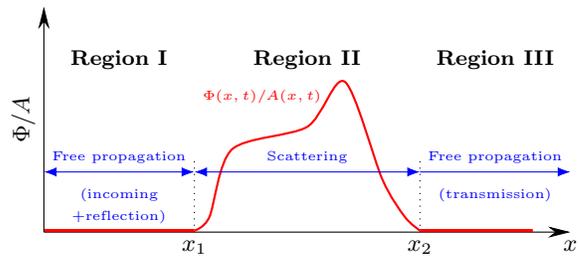

\bigskip

The technical difficulty in solving the problem described above is that regions I and III are in principle semi-infinite. In other words, focusing only on a finite interval encompassing the interaction region, we face an open quantum system, which is in a strong connection with its surroundings: e.g., the transmitted wave propagates outside the region of interest. For sinusoidal excitations, this problem can be handled by using Floquet's theory \cite{F883,SPF15}, but for pulsed excitations a different approach is needed.

For localized, static potentials a method based on non-equilibrium Green's functions (NEGF) is proven to be very effective for the description of transport processes in nanoscale devices, where semi-infinite leads are assumed to be connected to the region of interest. Without going into details (that can be found e.g. in Ref.~\cite{D95}), the main idea is to use analytic results for the semi-infinite regions, in which the potential is zero (and quantum mechanical waves propagate freely). For a given input energy $E$, the scattering matrix connecting the two leads (which is in a direct connection with the transmission probability) is shown to be proportional to the retarded Green's function (matrix) $G^R$ of the problem \cite{FL81}. Using effective mass approximation and spatial discretization, one faces the problem of inverting an infinite matrix to obtain $G^R,$ which is essentially $(H-E)^{-1}.$ However, since the problem is "simple" in the leads (free propagation), analytic results exist for their contribution. In this way, the matrix one has to invert will be finite, in fact only the matrix elements of the Hamiltonian that connect the interaction area to the leads has to be modified, in order to take the effects of the leads into account. In this way, a numerically exact solution of the scattering problem can be obtained. This method will be used in Subsec.~\ref{lorisubsec} in order to calculate the solution of static scattering problem as a reference.

For time dependent excitations, a time dependent version of the non-equilibrium Green's function approach (TDNEGF) is to be used. In the single electron picture, this approach can be turned into direct numerical method \cite{G14,F15}. Practically, the difficulty to be handled is allowing the disturbance caused by the excitation to leave the interaction region undisturbed. For a general time-dependent excitation, the behaviour of the wave function at the boundaries of the interaction region is a result of the time evolution, thus cannot be known in advance. The numerical version of the single particle TDNEGF method allows us to modify the time derivative of the wave function at the boundaries of the interaction region to mimic the semi-infinite leads, similarly to the static case. The numerical price of the ideally transparent boundary conditions is that the corrections at the boundaries involve time integrals that keep track the probability current that has flown out of the interaction region. This approach will be used to calculate the consequences of time-dependent excitations.

\section{Results}
\label{resultsec}
\subsection{Switch-on effects}
\label{lorisubsec}
Let us start considering the case when an external potential is being suddenly switched on. This can serve as a model for gate voltage induced transients in metallic solids. Concretely, we assume that $A=0$ in Eq.~(\ref{Ham}), while the scalar potential is given by
\begin{equation}
\Phi(x,t)= \varphi(x) \chi(t)
\label{switchpot}
\end{equation}
with
\begin{equation}
\varphi(x)=
\begin{cases}
\varphi_{max}\sin^2{\left(\frac{5\pi}{L}x\right)} & x \in \left[0,\frac{L}{10}\right], \\
\varphi_{max} & x \in \left[\frac{L}{10},\frac{9 L}{10}\right], \\
\varphi_{max}\left\lbrace 1-\sin^2{\left[\frac{5\pi}{L}x-4.5\pi\right]}\right\rbrace & x \in \left[\frac{9 L}{10},L \right].\\
0 & \text{otherwise.}
\end{cases}
\label{spatial}
\end{equation}
The time dependent part of the scalar potential (i.e., $\chi(t)$) represents a fast (but not instantaneous) switch on with a $\sin^2$ envelope for Figs.~2 and 3, and also a similarly decaying switch off for Fig.~4. The actual behaviour of $\chi(t)$ is shown by the top panels of Figs.~2 and 4.

In order to be able to compare the current results to the case of a ponderomotive potential to be discussed in the next subsection, we consider $\Phi(x,t)$ above as a potential barrier for the electrons. The appearance of the potential hill changes the initially uniform electron density. For a given input energy $E(k)$ and a potential with a height that corresponds to a classically forbidden region for $E(k)$ (and has an extension that renders quantum mechanical tunneling practically impossible), the transmission will be zero on the long time limit. The slope of the potential forces the electron density out of the interaction region, leading to transient wave packets travelling in both the reflected and transmitted directions. On the transmission side of the potential, because of the impenetrable barrier, there will be no input current, thus after the disappearance of the transients, the probability density will be zero. For the other side of the potential barrier, however, there is a continuous input travelling towards the barrier. The incoming plane wave is totally reflected by the potential, and the interference of oppositely travelling waves creates a regular pattern that appears to propagate in the negative direction with a velocity determined by the reflected transients.

\begin{figure}
\begin{center}
\includegraphics[width=\linewidth]{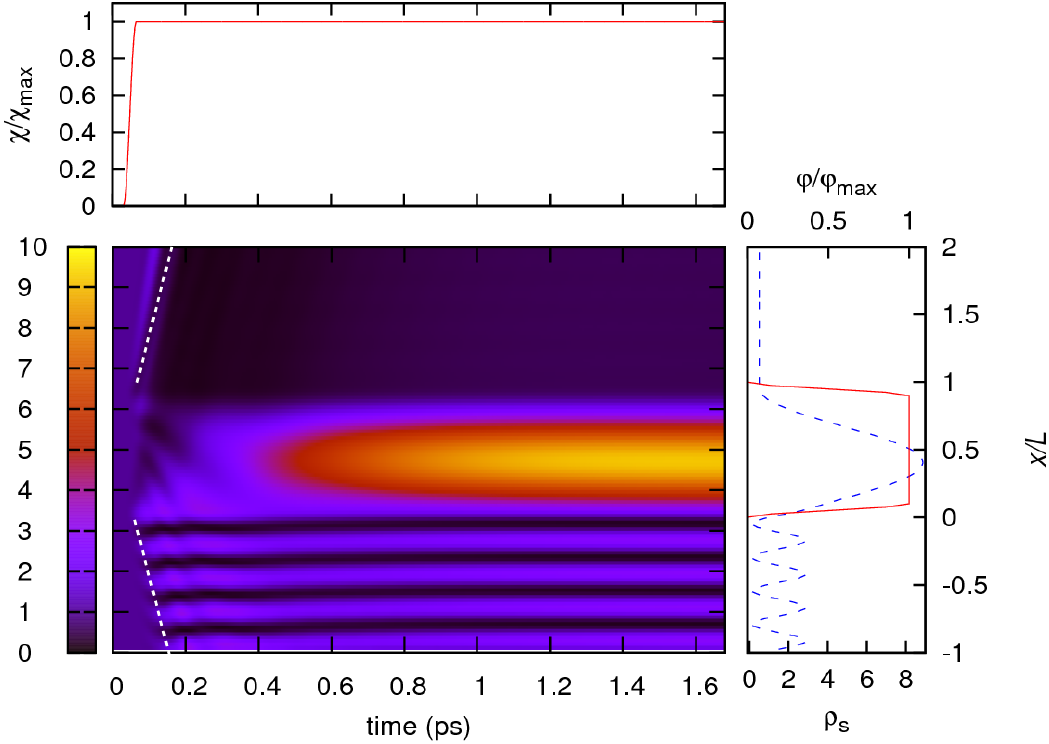}
\caption{
Central panel: Space and time dependence of the electron density as induced by the time dependent potential shown on the top. The right panel shows the steady state electron density (blue dashed line) and the spatial dependence of the potential (solid red line). The maximum of the potential was chosen so that the steady state transmission probability is $T=1/2$ for the initial electron energy of $E(k)=54$ meV. The dashed white lines guide the eyes, they represent a motion with a constant band velocity.
}
\label{timedepfig}
\end{center}
\end{figure}

The steady state interference pattern can be calculated using the static NEGF method sketched in the previous section, allowing us to calculate a transmission probability $T,$ which is constant both in space and time. Typical transient dynamics for the case when $T$ is not very close to zero is shown in Fig.~2. As we can see, the main effects we mentioned for an impenetrable potential above, also appear in this case. We can clearly identify the decrease of the electron density on the transmission side of the barrier, and the gradual appearance of the interference pattern on the reflection side of the barrier is also visible. The time scale of these effects can be calculated by investigating the slopes of the dashed white lines in Fig.~2, which turn out to be very close to the band velocity $v_k=\frac{1}{\hbar}\frac{\partial E(k)}{\partial k}$ (which is equal to $\frac{\hbar k}{m}$ for the parabolic bands we consider).

There are, however, slower mechanisms that finally lead to the steady state solution. E.g., the high electron density peak at the reflection side of the potential is formed on a longer time scale. The intuitive reason for the appearance of this peak is that since the incoming plane wave cannot be perfectly transmitted through the potential, its reflected part interferes with its own continuously arriving input part and results in an accumulation of the electron density. The regular density pattern on the reflection side of the potential reaches its steady state structure also on a longer time scale: although the primary structure becomes visible practically as soon as the first reflected wave packet arrives, the heights of the interference peaks converge to their final, steady state value on a much longer time scale. That is, while the part of the problem that can be understood by merely classical considerations (i.e., the expulsion of the electron density from the interaction area) has a time scale that is determined by the band velocity, interference related effects need more time to be built up. This can be seen in Fig.~3, where the difference
\begin{equation}
D(t)=\int \left[\rho_s(x) - \rho(x,t)\right] dx
\label{dist}
\end{equation}
is shown. In the equation above, the electron densities have the usual definition of modulus square of the wave function, and $\rho_s(x)$ corresponds to the steady state density that is constant in time, while $\rho(x,t)$ is calculated from the actual, proper wave function that includes transients. The integration domain is the whole interaction region. As we can see, for all curves in Fig.~3, $D(t)$ approaches zero for increasing values of $t,$ which is natural since the transients are expected to leave the interaction area in the long time limit. Additionally, we can also see an initial, fast decrease of all $D(t)$ curves, which corresponds to the effects emphasized by the white dashed lines in Fig.~2. The time instant when the forward scattered wave packets leave the interaction region correspond to pronounced drops (indicated by the arrows in Fig.~3) of the function $D(t).$ Clearly, the behaviours of the curves depend on the input electron energies: when the corresponding band velocity is larger, convergence towards the steady state electron density is faster. This could be expected for the initial part of the time evolution, but according to our results, it also holds for the final, slower, interference-dominated part of the dynamics.

\begin{figure}
\begin{center}
\includegraphics[width=\linewidth]{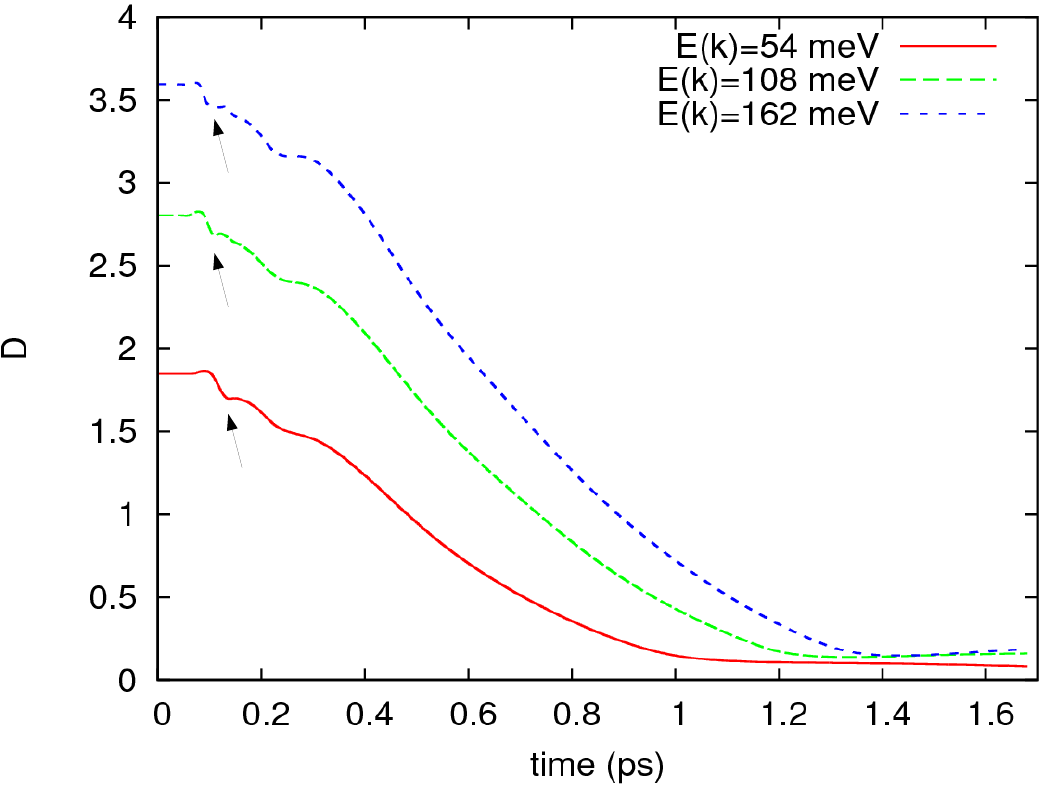}
\caption{
The distance $D$ defined by Eq.~(\ref{dist}) between the steady state solution and the actual electron density for different electron energies $E(k).$ For the sake of easy comparison, the steady state transmission probability is $T=1/2$ for all curves. The input electron energies are indicated by the legend. The arrows correspond to time instants when the first pronounced wave packets leave the interaction region on the transmission side of the potential (e.g., see the upper dashed white line in Fig.~2.)
}
\label{timedepfig}
\end{center}
\end{figure}

Note that the results shown by the figures are representative, but unavoidably particular examples. Although the qualitative picture is the same as discussed above, different parameter settings influence the details of the process of convergence towards the final, steady state solution. Generally, the effect of the sudden emergence of the potential barrier results in the broadening of the initially infinitely narrow quasimomentum distribution. $\Psi_0,$ which corresponds to a well defined value of $k=k_0,$ will be transferred into a superposition of plane waves with different qausimomenta. This distribution has two dominant peaks at $\pm k_0,$ which determine the fast part of the time evolution, but the convergence is not complete until those parts of the wave packet that belong to smaller quasimomenta slowly leave the interaction area. Additionally, when defining a characteristic time for the convergence (e.g., by requiring $D(t)$ to decrease below a predefined limit), we always have to specify the interval on which we would like to observe convergence. Clearly, the further away the observation point from the potential barrier is, the more time is needed for the transients to disappear.

\begin{figure}
\begin{center}
\includegraphics[width=\linewidth]{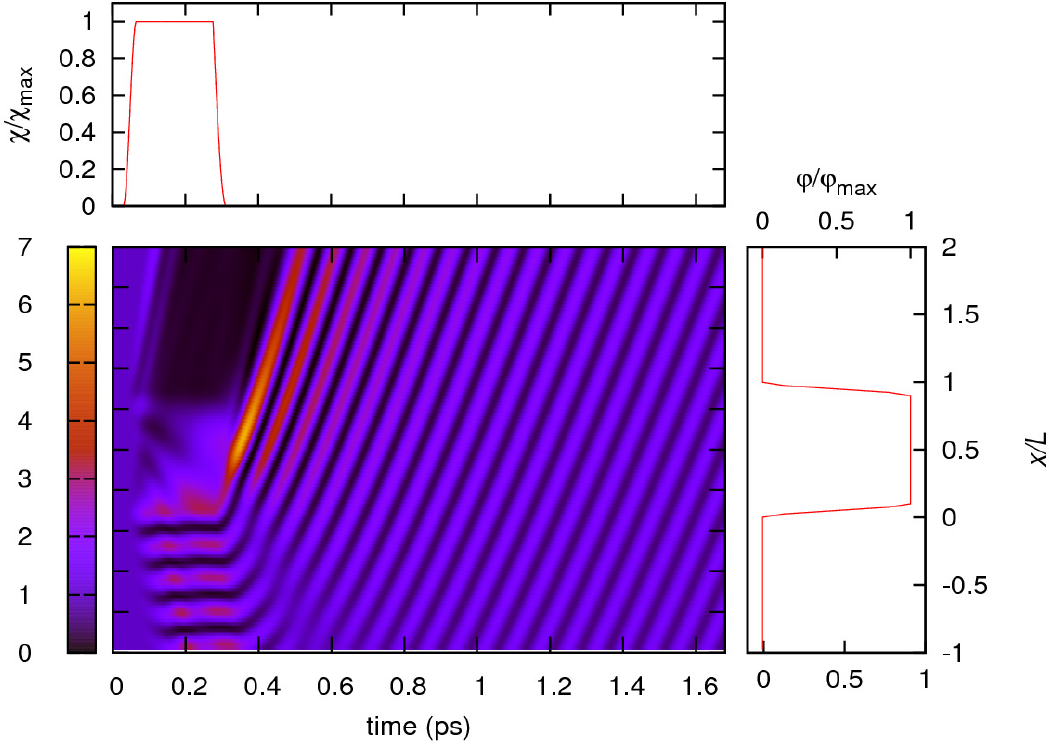}
\caption{
Central panel: Space and time dependence of the electron density as induced by the time dependent potential shown on the top. Right panel: The spatial dependence of the potential (its maximum is the same as in Fig.~2.)
}
\label{timedepfig}
\end{center}
\end{figure}

\bigskip

It is also interesting to ask what happens when the suddenly emerging potential stays constant in time for a finite interval, then it becomes zero again. This corresponds to the case of switching on and off of a gate voltage that influences the electron dynamics. According to the example shown in Fig.~4, the most dominant effect in this case is the generation of wave packets that propagate in the same direction as the initial plane wave. In view of the previous results, we can interpret the first maximum as the propagation of the electron density that was pushed out (in the positive direction) of the interaction region by the emergence of the potential. The minimum following this peak is a consequence of the decrease of the transmission because of the potential barrier. The following peaks are new in the sense that they are not consequences of switching on the potential. Instead, they mean essentially the propagation of the pronounced density maximum that can be seen in Fig.~2, as well as that of the whole interference pattern on the reflection side of the potential. (Intuitively: the amassed probability density is released when the potential barrier vanishes.) During the propagation, weaker interference maxima merge, and essentially a double peaked wave packet remains. The minimum between the two peaks is a fingerprint of the time window when the potential was on, and the transmission was not unity.

Clearly, the result shown in Fig.~4 is not completely general. Depending on the time difference between switching on and off the potential, the time evolution of the electron density can be qualitatively different. (E.g., when the time between switching on and off the potential is considerably larger than the one shown in the figure, two separated wave packet families arise.) However, the parameters we used for Fig.~4 can mimic the ponderomotive potential of a localized laser excitation, and thus can help interpreting the results of the next subsection.

\subsection{Laser pulse induced localized excitations}
\label{istvansubsec}
Now we describe laser-matter interaction in the velocity gauge and assume that $\Phi=0,$ while the space and time dependence of the $x$ component of the vector potential is given by
\begin{equation}
A(x,t) = A_0 \sin^2 \left( \frac{\pi}{L} x \right) \sin^2\left( \frac{\pi}{\tau} t \right) \sin \left( \omega_0 t \right)	
\end{equation}
if $x \in [0,L]$ and $t \in [0,\tau]$, otherwise $ A(x,t)=0$. In the simulations we consider the central wavelength of the laser to be $ \lambda_0 = \frac{2 \pi c}{\omega_0} = 800$ nm, and assume a pulse duration of $ \tau = 26,7$ fs (meaning 10 complete oscillations). The electric field can be calculated as the negative of the time derivative of $A(x,t)$, and we set the amplitude of the external electric laser field to $ F_0 = 1$GV/m.

It is important to emphasize that using dipole approximation, i.e., assuming that the vector potential has no spatial dependence, the time evolution induced by the Hamiltonian (\ref{Ham}) would be trivial for an initial plane wave. Since plane waves are eigenstates of $H(t)$ for $\frac{\partial A}{\partial x}=0,$ there will be no transitions between different plane wave states in this case. This is clearly related to the well-known fact that free electrons do not gain energy during the interaction with a single electromagnetic mode. That is, the localization of the excitation is crucial from the viewpoint of nontrivial time evolution (at least for a single band model).

Technically, for the calculation of the response of the electrons to the localized laser pulse, we found that instead of the TDNEGF approach, it is more efficient to use a Fourier transform-based method. In more details, we consider a relatively large computational box for the region II shown in Fig.~1. In this way the excitation-induced wave packets have no time to reach the boundaries of the computational box until the laser pulse is over. Then we extend the computational box by roughly ten times of its original size, apply periodic boundary conditions and use spatial Fourier transform to obtain $k$-resolved states, the time evolution of which are trivial, i.e., plane wave-like. Using a sufficiently large extended computational box, the effectiveness of the fast Fourier transform algorithm leads to a fast numerical method that provides an adequate description of the problem until the transients leave the (original) region II.

\begin{figure}
\begin{center}
\includegraphics[width=\linewidth]{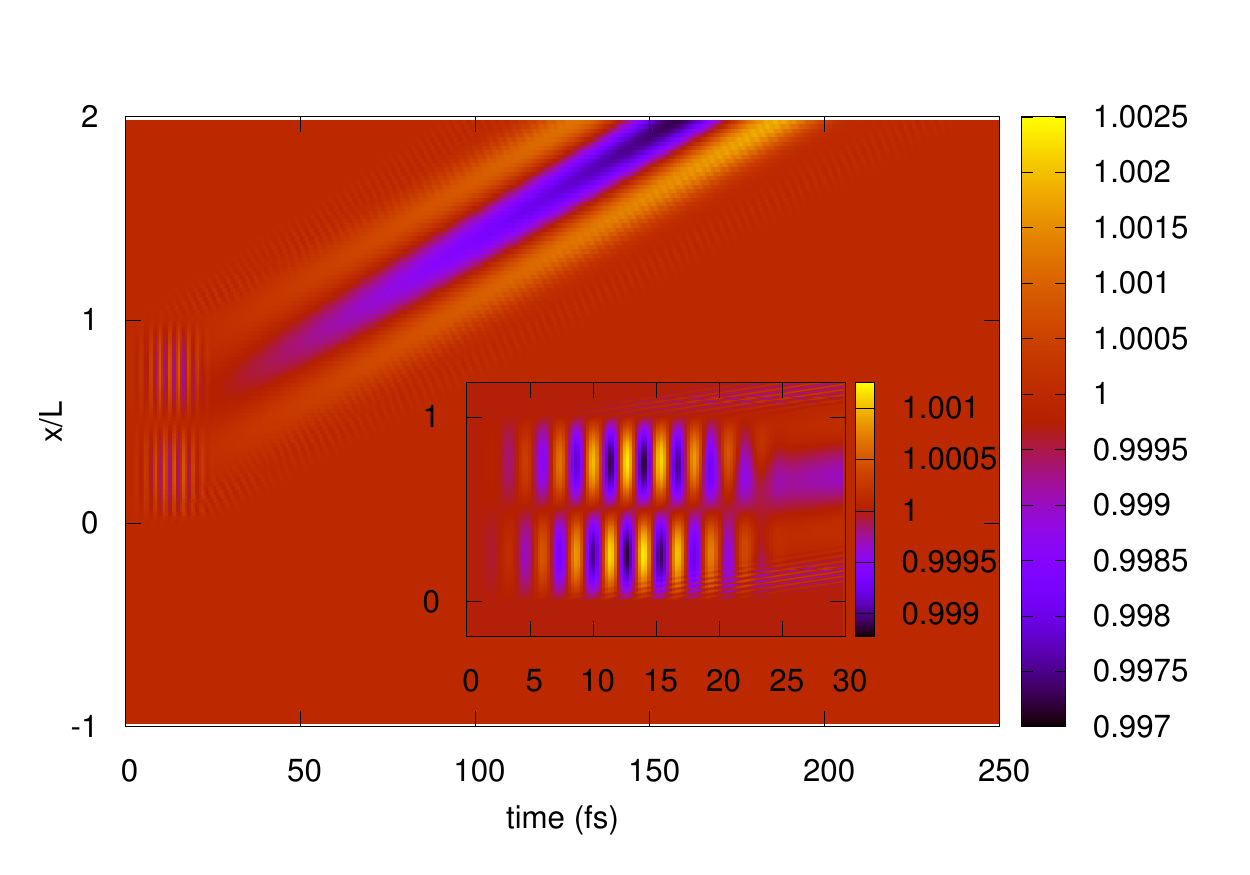}
\caption{
Space and time dependent electron density. The inset zooms on the domain on which the excitation is nonzero. Parameters are: $ F_0 = 1$GV/m, $L = 160$nm, $ \tau = 26,7$fs, $ \lambda_0 = 800$nm.
}
\label{timedepfig}
\end{center}
\end{figure}

Fig.~5 shows the typical space and time dependence of the electron density. As we can see, during the presence of the laser signal, the electric field induces oscillatory electron motion in the interaction area. Depending on its sign, the external field moves the electrons towards the positive or negative $x$ direction. The corresponding density intensifications and attenuations leave the interaction area as narrow wave packets, the periodicity of which is the same as that of the laser carrier oscillations (see the inset in Fig.~5.). However, the amplitude of these oscillations strongly decreases as we increase the size of the interaction area, which is directly related to the fact mentioned in the first paragraph of this subsection. This effect is so strong, that for traditional focusing that corresponds to spot sizes being a few times larger than the wavelength, oscillations with the laser carrier frequency are practically unseen (unless we calculate Fourier transforms, see below). In order to show these oscillations on the figure, interaction lengths below the wavelength of the exciting laser pulse were considered, which can be achieved e.g., by nanolocalized fields (see e.g. \cite{SS11,FM15}).

\begin{figure}
\begin{center}
\includegraphics[width=\linewidth]{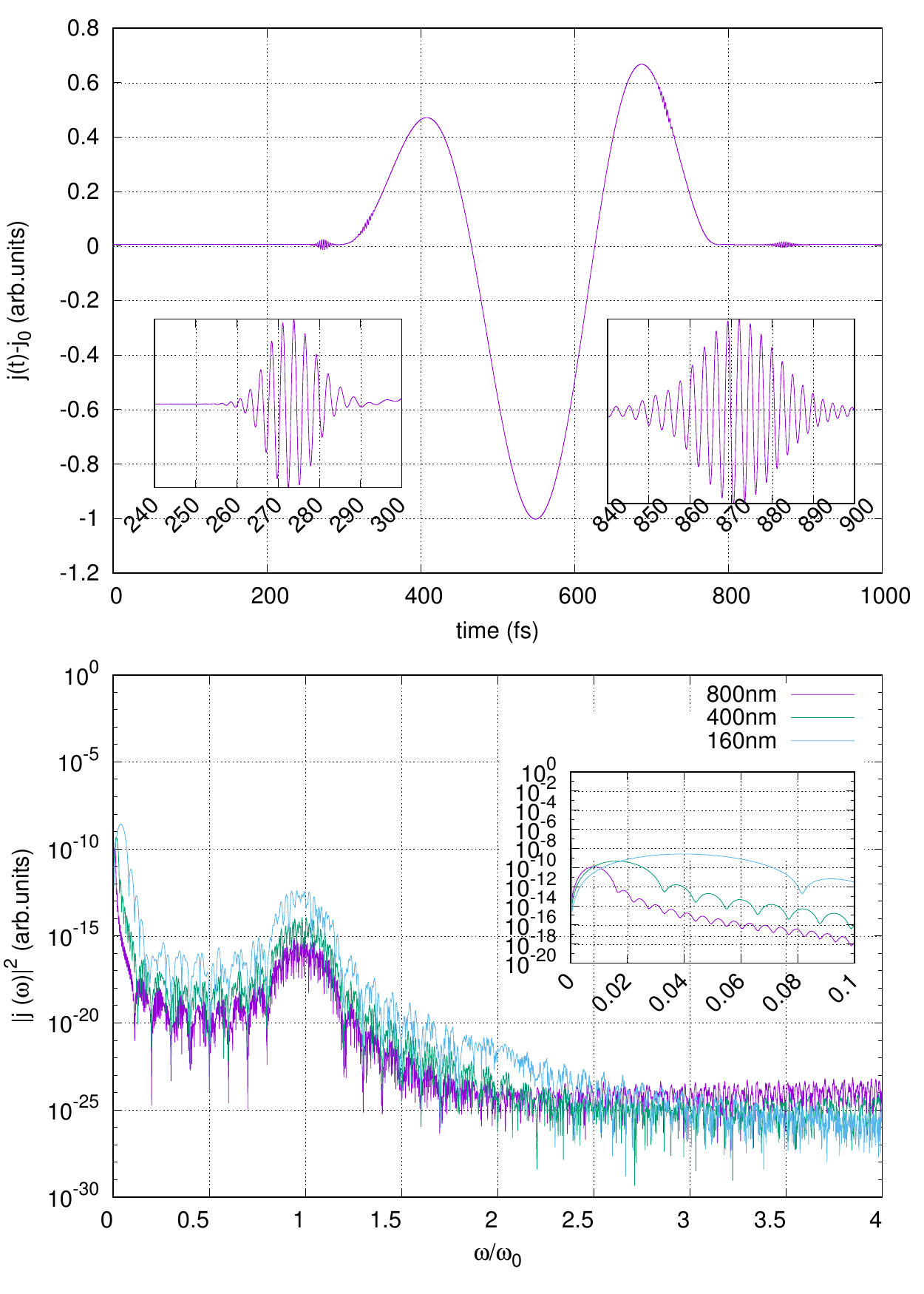}
\caption{
Top panel: typical time dependent current density calculated at the point $x=2 L$. Bottom panel: power spectra of the time dependent signals for the sizes of the interaction area ($L$) indicated by the legend. Parameters: $ F_0 = 1$ GV/m, $ \tau = 26,7$fs, $ \lambda_0 = 800$nm.
}
\label{timedepfig}
\end{center}
\end{figure}

On the other hand, the exact limit of spatially independent excitations can practically never be reached, since the focal spot of the exciting laser pulse is obviously finite. This localization leads to the most pronounced wave packet structure shown in Fig.~5. The similarity of these double peaked density waves to the ones that appear for non-oscillating potentials suggests a common interpretation. Indeed, besides the mere visual correspondence, the space dependent ponderomotive potential of the laser field
\begin{equation}
U_p(x,t)=\frac{e^2\mathcal{E}_0^2(x,t)}{4m\omega_0^2}
\end{equation}
can play the role of the potential that were considered in the previous subsection. (Here $\mathcal{E}_0(x,t)$ denotes the slowly varying local amplitude of the electric field strength, the time dependence of which stems from the change of the envelope of the laser field.) In other words, the finite duration of the excitation means a ponderomotive potential that is switched on and off on the time scale of the duration of the laser pulse. This potential produces the double peaked density waves seen in Fig.~5, similarly to the case discussed in the previous subsection.

Let us recall, however, that the concept of the ponderomotive potential involves averaging over an oscillation period. That is, although $U_p$-related effects are seen for all parameter ranges that we investigated, they are expected to be stronger when -- in classical terms -- the electron spends more time in the interaction zone. The analysis of the current density $j$ measured at the transmission side boundary of the interaction area completely supports this qualitative picture.

Fig.~6 shows the time dependence of $j$ as well as its frequency spectrum for three different sizes of the interaction area. (Note that effects related to localized excitation (in the sense of quantum confinement) has recently been investigated in Ref.~\cite{MAAB17}). The general feature of $j(t)$ shown in the figure is that it contains three different signals: two relatively weak, fast oscillating wave packets, which compass a more slowly oscillating part with a larger amplitude. Comparing with two-dimensional plots as shown by Fig.~5 for the particle density, the weaker signals can be identified as consequences of the direct laser-driven charge oscillations: they are born during the interaction with the laser field, at both sides of the interaction area (where the gradient of the field is the largest), while the more slowly varying part is $U_p$-related. This is also seen in the power spectra $|j(\omega)|^2,$ which have two distinct peaks corresponding to the time signals described above. Note that the central frequency of the laser, $\omega_0,$ is conveniently compared to $\omega(k)=E(k)/\hbar$ in order to identify different regimes, but the frequency $\omega(k)$ is practically absent from the spectra. The reason for this is that the current density involves the product of the wave function and its complex conjugate, thus it is not sensitive to the quantum mechanical phase of the wave function.

\begin{figure}
\begin{center}
\includegraphics[width=\linewidth]{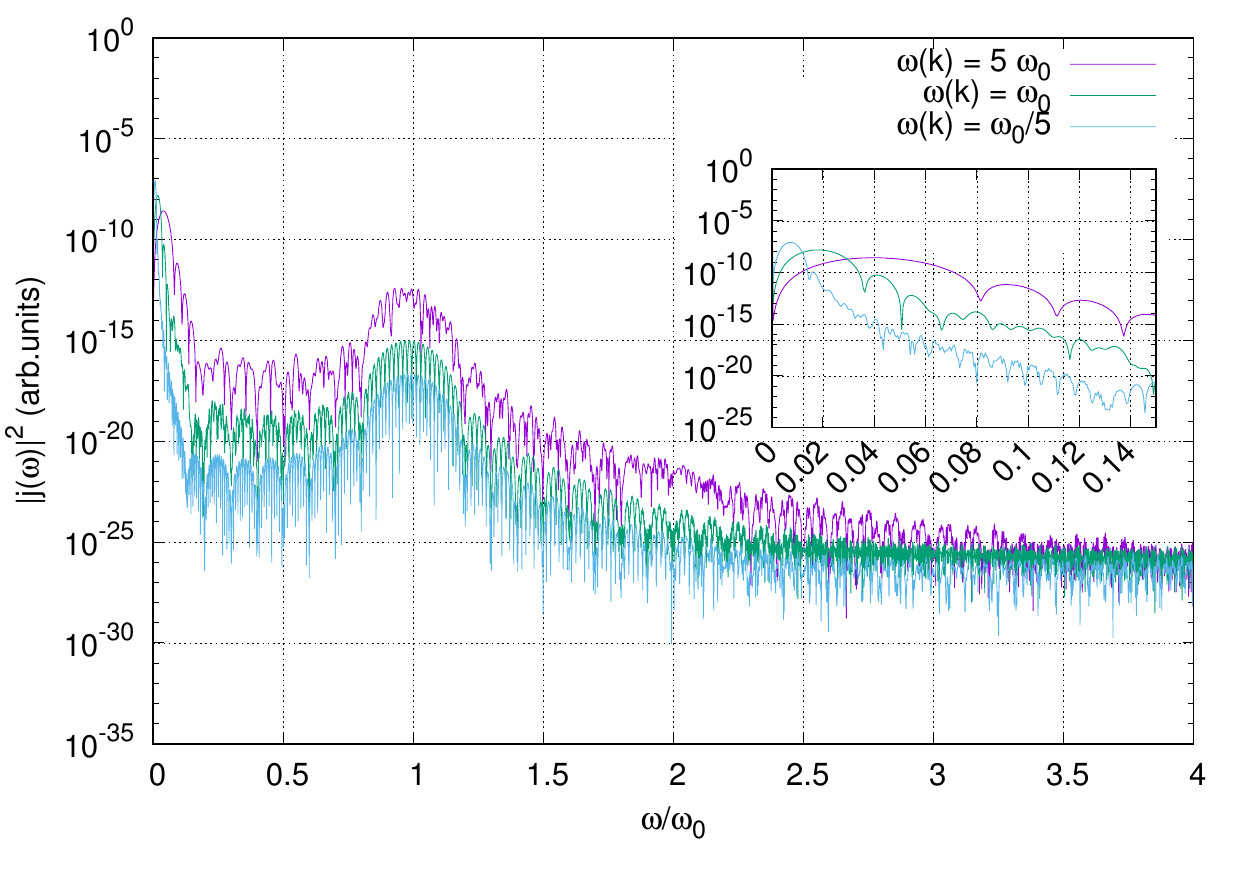}
\caption{
Power spectra $|j(\omega)|^2$ calculated at the point $x=2L$ for different initial electron energies $E(k)=\hbar\omega(k)$. Parameters are the same as in Figs.~5.
}
\label{timedepfig}
\end{center}
\end{figure}

The relative weights of the $\omega_0$ and $U_p$-related peaks in the spectra have opposite dependences on the parameters we consider. As we can see in Figs.~6 and 7, oscillations with the frequency of $\omega_0$ get weaker for larger interaction areas as well as for lower electron energies $E(k).$ This is in agreement with the considerations in the first paragraph of this subsection: in classical terms again, the larger the effective interaction area is (the more time the electron spends in the laser field), the more appropriate the dipole approximation is, and the weaker laser-induced effects are expected. Although the ponderomotive potential is also slightly less effective for larger interaction areas (since the gradient of the field envelope is smaller), this effect is much weaker than the decrease of signal at $\omega_0.$ Therefore, as the interaction area increases, $U_p$-related effects become dominant.

Additionally, slow electrons experience more laser cycles, which, as Fig.~7 shows, leads to stronger low-frequency peaks in $|j(\omega)|^2.$ Moreover, while the peaks around $\omega_0$ in the spectra shown in Figs.~6 and 7 have practically the same widths, narrower interaction areas and higher electron energies lead to shorter $U_p$-related signals in time domain, and consequently the corresponding frequency domain peaks will be broader. This is in complete agreement with the intuitive picture we can associate to $U_p$ as a potential that is being switched on and off on the timescale of the pulse duration.

\section{Summary}
\label{summarysubsec}
Conduction band electrons were investigated in localized, time-dependent external fields. For switching on a potential that represents a gate voltage, it was shown that the convergence towards the steady state solution is determined by two time scales, a faster one that corresponds to the band velocity, and a slower one that describes the build up of the steady state interference pattern. For a laser pulse, the mechanisms behind the behaviour of the electrons are: laser driven oscillations and the effects of the ponderomotive potential. By investigating the frequency spectra of the time dependent current induced by the laser pulse, we have shown and explained that for large interaction area and low electron energies, the time evolution is dominated by the low frequency, few-cycle oscillations that are induced by the ponderomotive potential. Electronic answer at the same frequency as the exciting field is expected to be appear for spatially narrow excitations and high energy electrons. Our results are relevant for using metallic targets as detectors for the parameters of the laser pulse.

\bigskip


The ELI-ALPS project (Grants No.  GOP-1.1.1-12/B-2012-000  and  No.  GINOP-2.3.6-15-2015-00001)  is  supported  by  the  European  Union  and  co-financed by the European Regional Development Fund.
Our work was also supported by the European Social Fund (EFOP-3.6.2-16-2017-00005).



\end{document}